\documentclass[aps,prl,preprint,tightenlines,superscriptaddress,showpacs,byrevtex]{revtex4}
\usepackage{graphicx}
\usepackage{xspace}
\usepackage{epsfig}
\usepackage{multirow}

\newcommand{\de}{\ensuremath{\Delta E}\xspace}

\newcommand{\acp}{\ensuremath{\mathcal{A}_{CP}}\xspace}

\newcommand{\bb}{\ensuremath{B \overline{B}}\xspace}

\def\myspecial#1{}                   

\def\SIGMA{\mbox{$5.4$}}

\def\BR{
  \mbox{$(1.1 \pm 0.3({\rm stat.}) \pm 0.1({\rm syst.})
  )\times 10^{-6}$}}

\def\bb{\mbox{$B\overline{B}$}}
\def\qq{\mbox{$q\overline{q}$}}
\def\mbc{\mbox{$M_{\rm bc}$}}
\def\de{\mbox{$\Delta E$}}
\def\acp{\mbox{${\cal A}_{CP}$}}

\def\bz{\mbox{$B^0$}}

\def\bp{\mbox{$B^+$}}

\def\dz{\mbox{$D^0$}}
\def\dzb{\mbox{$\overline{D}{}^0$}}

\def\kppim{\mbox{$K^+ \pi^-$}}

\def\pizpiz{\mbox{$\pi^0 \pi^0$}}

\def\dzbpip{\mbox{$\dzb(\to\kppim\pi^0)\pi^+$}}

\begin{document}

\myspecial{!userdict begin /bop-hook{gsave 300 50 translate 5 rotate
    /Times-Roman findfont 18 scalefont setfont
    0 0 moveto 0.70 setgray
    (\mySpecialText)
    show grestore}def end}


\vspace*{-3\baselineskip}
\resizebox{!}{3cm}{\includegraphics{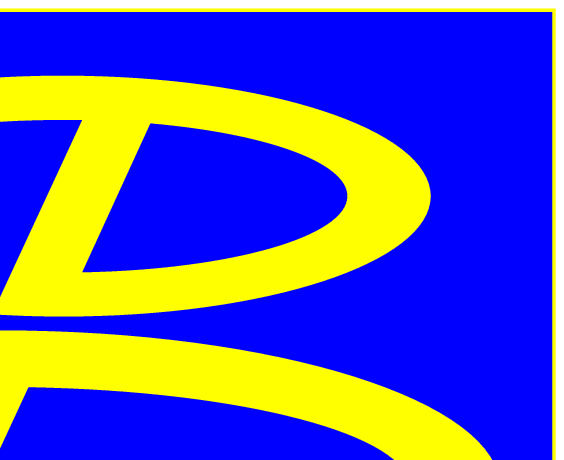}}

\preprint{\vbox{ \hbox{   }
                 \hbox{   }
                 \hbox{   }
                \hbox{BELLE-CONF-0677}
}}

\title{\quad\\[0.5cm] 
Improved measurement of $B^0 \to \pi^0 \pi^0$}

\affiliation{Budker Institute of Nuclear Physics, Novosibirsk}
\affiliation{Chiba University, Chiba}
\affiliation{Chonnam National University, Kwangju}
\affiliation{University of Cincinnati, Cincinnati, Ohio 45221}
\affiliation{University of Frankfurt, Frankfurt}
\affiliation{The Graduate University for Advanced Studies, Hayama} 
\affiliation{Gyeongsang National University, Chinju}
\affiliation{University of Hawaii, Honolulu, Hawaii 96822}
\affiliation{High Energy Accelerator Research Organization (KEK), Tsukuba}
\affiliation{Hiroshima Institute of Technology, Hiroshima}
\affiliation{University of Illinois at Urbana-Champaign, Urbana, Illinois 61801}
\affiliation{Institute of High Energy Physics, Chinese Academy of Sciences, Beijing}
\affiliation{Institute of High Energy Physics, Vienna}
\affiliation{Institute of High Energy Physics, Protvino}
\affiliation{Institute for Theoretical and Experimental Physics, Moscow}
\affiliation{J. Stefan Institute, Ljubljana}
\affiliation{Kanagawa University, Yokohama}
\affiliation{Korea University, Seoul}
\affiliation{Kyoto University, Kyoto}
\affiliation{Kyungpook National University, Taegu}
\affiliation{Swiss Federal Institute of Technology of Lausanne, EPFL, Lausanne}
\affiliation{University of Ljubljana, Ljubljana}
\affiliation{University of Maribor, Maribor}
\affiliation{University of Melbourne, Victoria}
\affiliation{Nagoya University, Nagoya}
\affiliation{Nara Women's University, Nara}
\affiliation{National Central University, Chung-li}
\affiliation{National United University, Miao Li}
\affiliation{Department of Physics, National Taiwan University, Taipei}
\affiliation{H. Niewodniczanski Institute of Nuclear Physics, Krakow}
\affiliation{Nippon Dental University, Niigata}
\affiliation{Niigata University, Niigata}
\affiliation{University of Nova Gorica, Nova Gorica}
\affiliation{Osaka City University, Osaka}
\affiliation{Osaka University, Osaka}
\affiliation{Panjab University, Chandigarh}
\affiliation{Peking University, Beijing}
\affiliation{University of Pittsburgh, Pittsburgh, Pennsylvania 15260}
\affiliation{Princeton University, Princeton, New Jersey 08544}
\affiliation{RIKEN BNL Research Center, Upton, New York 11973}
\affiliation{Saga University, Saga}
\affiliation{University of Science and Technology of China, Hefei}
\affiliation{Seoul National University, Seoul}
\affiliation{Shinshu University, Nagano}
\affiliation{Sungkyunkwan University, Suwon}
\affiliation{University of Sydney, Sydney NSW}
\affiliation{Tata Institute of Fundamental Research, Bombay}
\affiliation{Toho University, Funabashi}
\affiliation{Tohoku Gakuin University, Tagajo}
\affiliation{Tohoku University, Sendai}
\affiliation{Department of Physics, University of Tokyo, Tokyo}
\affiliation{Tokyo Institute of Technology, Tokyo}
\affiliation{Tokyo Metropolitan University, Tokyo}
\affiliation{Tokyo University of Agriculture and Technology, Tokyo}
\affiliation{Toyama National College of Maritime Technology, Toyama}
\affiliation{University of Tsukuba, Tsukuba}
\affiliation{Virginia Polytechnic Institute and State University, Blacksburg, Virginia 24061}
\affiliation{Yonsei University, Seoul}
  \author{K.~Abe}\affiliation{High Energy Accelerator Research Organization (KEK), Tsukuba} 
  \author{K.~Abe}\affiliation{Tohoku Gakuin University, Tagajo} 
  \author{I.~Adachi}\affiliation{High Energy Accelerator Research Organization (KEK), Tsukuba} 
  \author{H.~Aihara}\affiliation{Department of Physics, University of Tokyo, Tokyo} 
  \author{D.~Anipko}\affiliation{Budker Institute of Nuclear Physics, Novosibirsk} 
  \author{K.~Aoki}\affiliation{Nagoya University, Nagoya} 
  \author{T.~Arakawa}\affiliation{Niigata University, Niigata} 
  \author{K.~Arinstein}\affiliation{Budker Institute of Nuclear Physics, Novosibirsk} 
  \author{Y.~Asano}\affiliation{University of Tsukuba, Tsukuba} 
  \author{T.~Aso}\affiliation{Toyama National College of Maritime Technology, Toyama} 
% ...
\collaboration{The Belle Collaboration}


\begin{abstract}
  We report an improved measurement of the decay $\bz\to\pizpiz$, using a
data sample of 535 $\times 10^6 B\bar{B}$ pairs collected
at the $\Upsilon(4S)$ resonance with the Belle detector at the 
KEKB asymmetric-energy $e^+ e^-$ collider.
  The measured branching fraction is ${\cal B}(\bz\to\pizpiz) = \BR$,
  with a significance of $\SIGMA$ standard deviations including
  systematic uncertainties.
  We also report the partial rate asymmetry: $\acp(\bz\to\pizpiz)$ = 0.44 $^{+0.73}_{-0.62}$(stat.)$^{+0.04}_{-0.06}$(syst.).
\end{abstract}

\pacs{11.30.Er, 12.15.Hh, 13.25.Hw, 14.40.Nd}
\maketitle

{\renewcommand{\thefootnote}{\fnsymbol{footnote}} 
\setcounter{footnote}{0}

\normalsize

\newpage
Measurements of the mixing-induced $CP$ violation parameter
$\rm{sin} 2\phi_1$~\cite{phi1_belle,phi1_babar} at $B$ factories
are in good agreement with the Kobayashi-Maskawa (KM)
mechanism~\cite{km}. 
To confirm this theory, one now has to measure
the other two angles of the unitarity triangle, $\phi_2$ and $\phi_3$.
%
One technique for measuring $\phi_2$ is to
study time-dependent $CP$ asymmetries
in $B^0 \to \pi^+\pi^-$ decay, where both Belle~\cite{phi2_belle} and 
BaBar~\cite{phi2_babar} recently 
reported the observation of mixing-induced $CP$ violation. 
Belle observed direct $CP$ violation while BaBar found no direct $CP$ violation.
The extraction of
$\phi_2$, however, is complicated by the presence of both tree and
penguin amplitudes, each with different weak phases. An isospin
analysis of the $\pi\pi$ system is necessary~\cite{isospin}, and
one essential ingredient is the branching
fraction for the decay $B^0 \to \pi^0\pi^0$.

QCD-based factorization predictions for ${\cal B}(B^0 \to
\pi^0\pi^0)$ are typically around or below $1 \times
10^{-6}$~\cite{pi0pi0_predictions}, but phenomenological models
incorporating large rescattering effects can accommodate larger
values~\cite{theory}. Observation for $B^0 \to \pi^0\pi^0$ was 
previously reported by Belle
with the value of $(2.3^{+0.4}_{-0.5}({\rm stat.})$$^{+0.2}_{-0.3}({\rm syst.}))\times 10^{-6}$ for the branching 
fraction~\cite{pi0pi0_Belle}.
If such a high value persists, an isospin
analysis for $\phi_2$ extraction would become feasible in the near
future. To complete the program, one would need to measure both the
$B^0$ and $\overline{B}{}^0$ decay rates, i.e. direct $CP$ violation.

In this paper we report the improved measurement of the decay
$\bz\to\pizpiz$. We also provide a measurement of the direct $CP$
violating asymmetry in this mode. The results are based on a 535 $\times 10^6
\bb$ pairs, collected  with the Belle detector at the KEKB asymmetric-energy
$e^+e^-$ (3.5 on 8~GeV) collider~\cite{KEKB}. 
 
The Belle detector is a large-solid-angle magnetic
spectrometer that
consists of a silicon vertex detector (SVD),
a 50-layer central drift chamber (CDC), an array of
aerogel threshold \v{C}erenkov counters (ACC),
a barrel-like arrangement of time-of-flight
scintillation counters (TOF), and an electromagnetic calorimeter
comprised of CsI(Tl) crystals (ECL) located inside
a super-conducting solenoid coil that provides a 1.5~T
magnetic field.  An iron flux-return located outside of
the coil is instrumented to detect $K_L^0$ mesons and to identify
muons (KLM).  The detector
is described in detail elsewhere~\cite{Belle}.
Two inner detector configurations were used. A 2.0 cm beampipe
and a 3-layer silicon vertex detector were used for the first sample
of 152 $\times 10^6 B\bar{B}$ pairs (Set I), while a 1.5 cm beampipe, a 4-layer
silicon detector and a small-cell inner drift chamber~\cite{SVD2}  
were used to record the remaining 383 $\times 10^6 B\bar{B}$ pairs (Set II).

Pairs of photons with invariant masses in the range $115 \ {\rm
MeV}/c^2 < m_{\gamma\gamma} < 152 \ {\rm MeV}/c^2$ are used to
form $\pi^0$ mesons and $\pi^0$ mass constraint is implemented; this corresponds to a window of $\pm
2.5\sigma$ about the nominal $\pi^0$ mass, where $\sigma$ denotes
the experimental resolution, approximately $8 \ {\rm MeV}/c^2$.
The measured energy of each photon in the laboratory frame is
required to be greater than $50 \ {\rm MeV}$ in the barrel region,
defined as $32^{\circ} < \theta_{\gamma} < 129^{\circ}$, and
greater than $100 \ {\rm MeV}$ in the end-cap regions, defined as
$17^{\circ} \le \theta_{\gamma} \le 32^{\circ}$ and $129^{\circ}
\le \theta_{\gamma} \le 150^{\circ}$, where $\theta_{\gamma}$
denotes the polar angle of the photon with respect to the positron beam
line. To further reduce the combinatorial background, $\pi^0$
candidates with small decay angles ($\cos\theta^* >0.95$) are
rejected, where $\theta^*$ is the angle between the $\pi^0$ boost
direction from the laboratory frame and one of its $\gamma$ daughters
in the $\pi^0$ rest frame.
 
Signal $B$ candidates are formed from pairs of $\pi^0$ mesons and
are identified by their beam energy constrained mass $\mbc =
\sqrt{E_{\rm beam}^{*2}/c^4 - p_B^{*2}/c^2}$ and energy difference $\de =
E_B^* - E_{\rm beam}^*$, where $E_{\rm beam}^*$ denotes the run-dependent beam
energy, $p_B^*$ and $E_B^*$ are the reconstructed momentum and energy of the 
$B$ candidates, all  in the $e^+e^-$ CM frame. 
We require $\mbc > 5.2 \ {\rm GeV}/c^2$ and $-0.45 \ {\rm GeV} < \de < 0.5 \ {\rm GeV}$.
The signal
efficiency is estimated using
GEANT-based~\cite{geant} Monte Carlo (MC) simulations. The resolution for signal is approximately $3.6 \ {\rm
MeV}/c^2$ in $\mbc$. The distribution in $\de$ is asymmetric due to
energy leakage from the CsI(Tl) crystals.

We consider background from other $B$ decays and from
$e^+e^-\to\qq$ ($q = u$, $d$, $s$, $c$) continuum processes. A
large generic MC sample shows that backgrounds from $b\to c$
decays are negligible. Among charmless $B$ decays, the only
significant background is $B^\pm \to \rho^\pm\pi^0$ with a
missing low momentum $\pi^\pm$. This background populates the negative
$\de$ region, and is taken into account in the signal extraction
described below.
 
The dominant background is due to the continuum processes. We use
event topology to discriminate signal events from this $\qq$
background, and follow the continuum rejection technique
from our previous publication~\cite{pi0pi0_Belle}. 
We combine a set of modified Fox-Wolfram moments \cite{fw} into a
Fisher discriminant~\cite{fisher}. 
A signal/background likelihood is formed, based on a
Monte Carlo (MC) simulation for signal and events in the $\mbc$ sideband region ($5.20 \ {\rm GeV}/c^2 < \mbc < 5.26 \ {\rm GeV}/c^2$) for the $\qq$ background,
 from the product of the probability density
functions (PDFs) for the Fisher discriminant and that for the cosine of the
angle between the $B$-meson flight direction and the positron beam. The 
continuum suppression is achieved by applying a requirement on a likelihood 
ratio ${\mathcal R}_{\rm sig} = {\mathcal L}_{\rm sig}/({\mathcal L}_{\rm sig} 
+ {\mathcal L}_{q \overline{q}})$, where
${\mathcal L}_{\rm sig}$ (${\mathcal L}_{q \overline{q}}$)
is the signal ($q \overline{q}$) likelihood.

Additional discrimination between signal and background can be
achieved by using the Belle standard algorithm for $b$-flavor
tagging~\cite{tagging}, which is also needed for the direct
$CP$ violation measurement. The flavor tagging procedure yields two
outputs: $q = \pm 1$, indicating the flavor of the other $B$ in the event,
and $r$, which takes values between 0 and 1 and
is a measure of the confidence that the $q$ determination is
correct. Events with a high value of $r$ are considered
well-tagged and are therefore unlikely to have originated from
continuum processes. For example, an event that contains a high momentum lepton
($r$ close to unity) is more likely to be a $B \overline B$ event
so a looser ${\cal R}_{\rm sig}$ requirement can be applied. We find
that there is no strong correlation between $r$ and any of the
topological variables used above to separate signal from the
continuum.
 
We divide the data into $r\ge 0.5$ and $r<0.5$ bins. The continuum
background is reduced by applying a selection requirement on
$\mathcal{R}_{\rm sig}$ for events in each $r$ region of Set I and Set II
according to the figure of merit (FOM). The FOM is defined as
$N_{\rm sig}^{\rm exp}/\sqrt{N_{\rm sig}^{\rm exp}+N_{\rm bg}^{\rm exp}}$, 
where $N_{\rm sig}^{\rm exp}$ and $N_{\rm bg}^{\rm exp}$ denote the expected 
signal,
assuming the branching fraction ${\cal B} = 2 \times 10^{-6}$,
and background yields obtained from $B^{\pm} \to \rho^{\pm} \pi^0$ MC and sideband data, respectively.
A typical requirement suppresses 98\% of the continuum background while
retaining 45\% of the signal.
  
\begin{figure}[htb]
\includegraphics[width=0.4\textwidth]{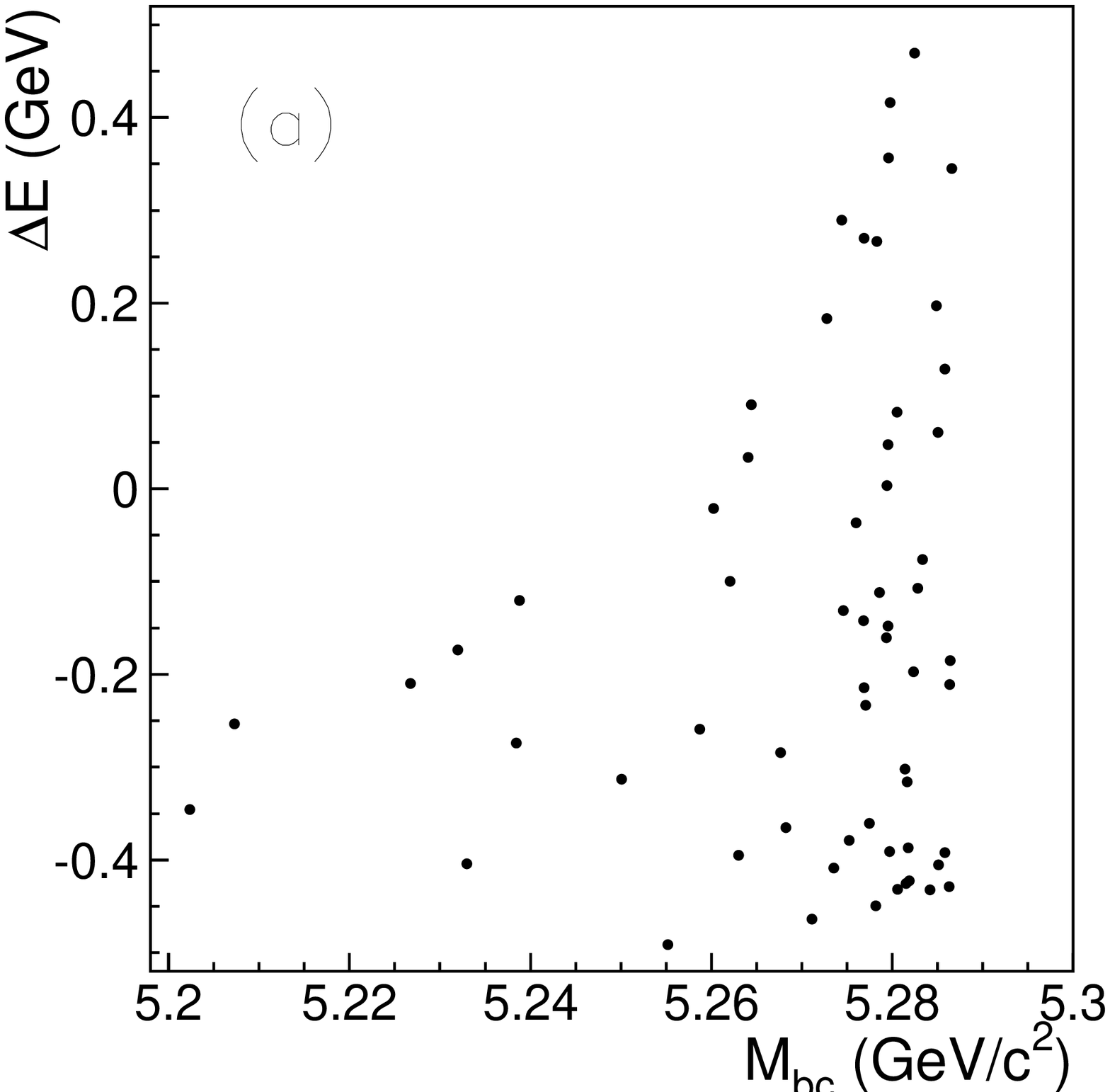}
\includegraphics[width=0.4\textwidth]{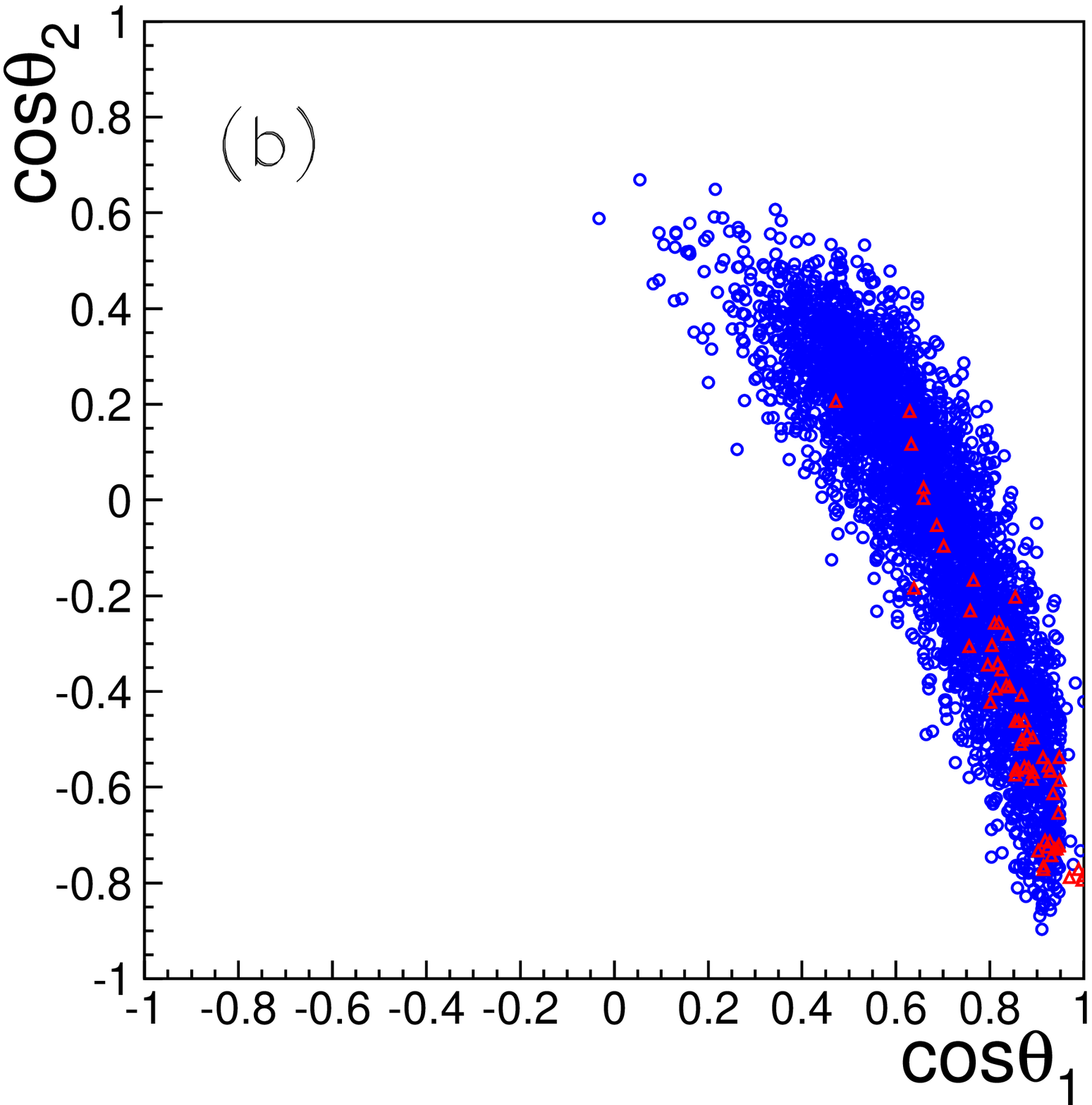}
   \caption{
    \label{fig:off} (a) The distribution of $\Delta E$ vs. $M_{\rm bc} $ for the off-time events; 
    (b) the distributions of $\cos \theta_2$ vs. $\cos \theta_1$
    for the on-time events (circle) and the off-time events (triangle).}
\end{figure}

In our previous publications \cite{pi0pi0_Belle, gg_Belle} 
we reported a special background originating from the overlap of a hadronic
continuum event and the residual calorimeter energies from an earlier QED 
scattering event.
This $e^+e^- \to \gamma \gamma$ background has a signature of back-to-back
photons in the CM frame, which can combine with two soft photons to form 
 $\pi^0 \pi^0$ candidates. The fake $\pi^0 \pi^0$ candidates tend to have very small total
CM momenta due to the characteristics of the 
back-to-back photons, resulting in a peaking behavior  at $M_{\rm bc}= 5.28 $
 GeV/$c^2$. Therefore the QED background is quite similar to the $B^0 \to \pi^0\pi^0$ 
signal and has to be considered in the signal extraction. The best way to
identify the  $e^+e^-\to \gamma\gamma$ events is to use the timing 
information of the ECL clusters, which are off-time for the QED background but  
only available in the latter 239 fb$^{-1}$ of data. For data in which the 
timing information is not available, we have to rely on other observables to 
distinguish candidates synchronized with the
trigger particles (on-time)  from the  off-time QED background. Assuming the 
distributions of the off-time QED background are the same for the first
253 fb$^{-1}$ and latter 239 fb$^{-1}$, the PDFs of the observables can be 
obtained using the data in the latter  239 fb$^{-1}$ sample, in which 
trigger timing information can be used to identify the on-time and off-time
events. 

Figure \ref{fig:off}(a) shows the 
$M_{\rm bc}$ - $\Delta E$ distribution for the off-time $\pi^0 \pi^0$ 
candidates in the latter  239 fb$^{-1}$ sample after passing all analysis 
requirements. These off-time $\pi^0 \pi^0$ events
 are located in the $M_{\rm bc}$ signal region but have no 
particular structure in $\Delta E$. Since the QED photons are close to the
$e^+$ and $e^-$ directions, the angle between the $\pi^0$ moving direction and 
the $z$ axis can be used to identify the QED background. The angular
 distributions of on-time and off-time candidates in the latter sample 
are shown in Fig.~\ref{fig:off}(b), where
$\theta_1$ and $\theta_2$ are the angles of higher and lower 
momentum $\pi^0$s, respectively. Compared to the on-time events,
the off-time candidates have a narrow distribution in
$(\cos\theta_1, \cos\theta_2)$-plane. 
The last variable that helps distinguish the QED background is missing energy,
defined as $E_{\rm miss}$ = $E^*_{\rm total}$ - 2  $E^*_{\rm beam}$, where
$E^*_{\rm total}$ is the total reconstructed energy in the CM frame. 
Since the QED background consists of two overlapping events, 
the missing energy tends to be larger than that of  the on-time events.
All these five observables are implemented in the fit.

The signal yields are extracted by applying unbinned five-dimensional
maximum likelihood (ML) fits to the ($M_{\rm bc}$,
$\Delta E$, $\cos \theta_1$, $\cos \theta_2$, $E_{\rm miss}$) distributions of 
the $B$ and $\overline B$ samples.
The likelihood is defined as
\begin{eqnarray}
\mathcal{L} & = & {\rm exp}\; (-\sum_{s,k,j} N_{s,k,j})
\prod_i (\sum_{s,k,j} N_{s,k,j} {\mathcal P}^i_{s,k,j}) \;\;\;
\end{eqnarray}
where
\begin{eqnarray} \mathcal{P}^i_{s,k,j} & = &
\frac{1}{2}[1-q^i \cdot \acp]P1_{s,k,j}(M^i_{\rm bc}, \Delta E^i) \times P2_{s,k,j}(\cos \theta^i_1, \cos \theta^i_2) \times P3_{s,k,j}(E^i_{\rm miss}).
\end{eqnarray}
The direct $CP$ asymmetry is defined as
\begin{eqnarray}
\acp \equiv \frac{N(\overline B \to \overline f)-N(B \to f)}
{N(\overline B \to \overline f)+N(B \to f)},
\end{eqnarray}
and $s$ indicates Set I or Set II, $k$ distinguishes events in the
$r<0.5$ or $r\ge 0.5$ bins, $j$ indicates the category of signal or background
contributions due to the $q\bar{q}$ continuum,  $B^\pm \to \rho^\pm\pi^0$ decay 
and off-time QED events.  $i$ is the identifier of the $i$-th
event, $P1_{s,k,j}(M_{\rm bc}, \Delta E)$ are the two-dimensional
probability density functions (PDFs) in $M_{\rm bc}$ and $\Delta E$,
$P2_{s,k,j}(\cos \theta_1, \cos \theta_2)$ are the two-dimensional PDFs in $\cos \theta_1$ and $\cos \theta_2$, $P3_{s,k,j}(E_{\rm miss})$ are the 
one-dimensional PDFs in $E_{\rm miss}$, $N_{s,k,j}$ is the number of events and 
$q^i$ indicates the $B$ meson flavor: $q^i$ = +1($-1$) for $B^0$ and $\overline{B}^0$.
The flavor of the  $B$ meson in the $B^0 \to \pi^0 \pi^0$ channel is not self-tagged and must be determined from the accompanying $B$
meson. To account for the effect of $B^0$-$\overline{B}^0$ mixing and 
imperfect tagging, the term $\acp$ for the signal in Eq. 2 has to be replaced by
$\acp (1-2\chi_d)(1-2w_k)$, where $\chi_d$ = 0.186 $\pm$ 0.004 \cite{PDG} is 
the time-integrated mixing parameter and $w_k$ is the wrong-tag fraction 
that depends on the value of $r$. The wrong-tag fractions are determined using 
a large sample of self-tagged 
$B^0 \to D^{*-}\pi^+, D^{*-}\rho^+$ and $D^{(*)-}l^+\nu$
events and their charge conjugates \cite{tagging}.

The $M_{\rm bc}$-$\Delta E$ PDFs for the signal and for the 
$B^+ \to \rho^+\pi^0$ background are taken
from smoothed two-dimensional histograms obtained from large MC
samples. For the signal PDF ($P1$), discrepancies between the peak positions and
resolutions in data and MC are calibrated using $\dz\to\pizpiz$ and
$\bp\to\dzbpip$ decays. The difference is caused by imperfect simulation
of the $\pi^0$ energy resolution, while the effect of the opening angle
distributions can be neglected. The invariant mass distribution for the $D^0$
is fitted with an empirical function for data and MC, and the
observed discrepancies in the peak position and width are converted to
the differences in the peak position and resolution for $\de$ in the
signal PDF. We require the $D^0$ decay products to lie in the same momentum
range as the $\pi^0$s from $B^0 \to \pi^0\pi^0$.
 To obtain the two-dimensional PDF, $P1(M_{\rm bc}$, $\Delta E)$, for the 
continuum background, we multiply a linear function for $\de$ with the ARGUS 
function~\cite{argus}  for $\mbc$.
The $M_{\rm bc}$-$\Delta E$ PDF for the off-time QED background is
modeled as a  smoothed two-dimensional histogram using the off-time candidates 
in the latter data set.

The $P2(\cos\theta_1,\cos\theta_2)$ and $P3(E_{\rm miss})$ PDFs are described by
 two-dimensional and one-dimensional 
smooth histograms, respectively. These PDFs are obtained using the on-time and off-time candidates with 
trigger timing information. Note that the same on-time PDFs are used for signals,  the $\rho^+\pi^0$ 
background and the continuum. In other words, these three components are distinguished based on
the $M_{\rm bc}$-$\Delta E$ distribution, while the off-time component is identified using all
5 variables.  In the fit, the shapes of the signal, off-time and
$B^+ \to \rho^+\pi^0$ PDFs are fixed and 
all other fit parameters are allowed to float. We check the modeling of the 
off-time QED background by comparing the results of the five-dimensional fit to
the data with timing information 
to the fit result using only the $M_{\rm bc}-\Delta E$ PDFs after removing 
the off-time candidates. With 3200 events, the obtained off-time yield from the 5-d fit is $79\pm 14$, consistent with 61 off-time candidates. 
Moreover, the obtained yields for signals, the $B^+\to \rho^+\pi^0$ background
 and the continuum are also consistent between
the two fits. 
 
\begin{figure}[htb]
\includegraphics[width=0.64\textwidth]{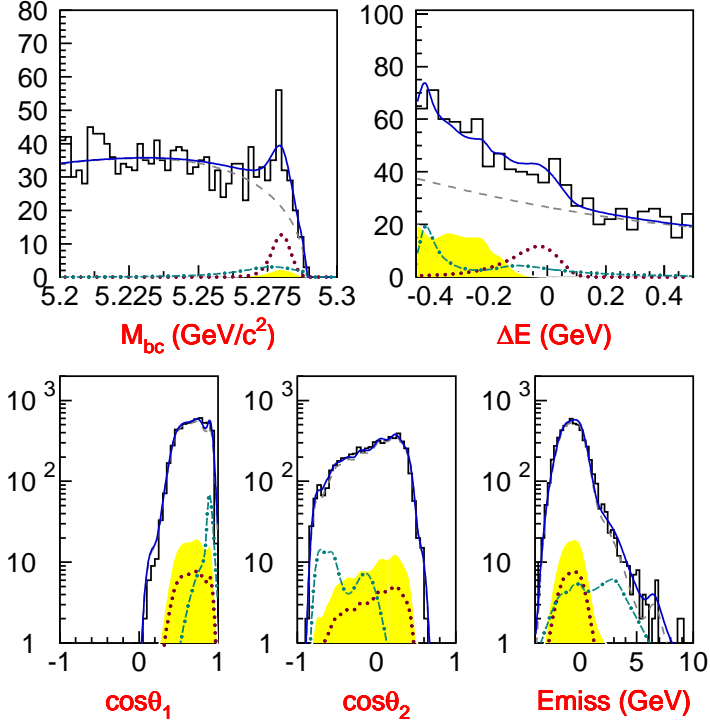}
  \caption{
    \label{fig:fit_result}
    Result of the fit described in the text.
    (Top-left) $\mbc$ projection for events that satisfy
    $-0.18 \ {\rm GeV} < \de < 0.06 \ {\rm GeV}$;
    (top-right) $\de$ projection for events that satisfy
    $5.27 \ {\rm GeV}/c^2 < \mbc < 5.29 \ {\rm GeV}/c^2$. The bottom plots from 
left to right are $\cos \theta_1$, $\cos \theta_2$ and $E_{\rm miss}$ 
projections without any $M_{\rm bc}$ and $\Delta E$ selection.
    The solid lines indicate the sum of all components,
    and the dotted, dashed, dot-dashed lines and hatched part
    represent the contributions from signal, continuum,
    off-time and $B^+ \to \rho^+\pi^0$ events, respectively.
  }
\end{figure}
 
We perform a 5-d fit to all the data assuming the same off-time
QED distributions for the former and latter data sets. Figure \ref{fig:fit_result}
shows the fit projections.
The obtained signal yield is $74.4^{+21.4}_{-19.7}$ with a statistical
significance (${\cal S}$) of $5.5 \sigma$, where ${\cal S}$ is defined as
${\cal S}=\sqrt{-2\ln({\cal L}_0/{\cal L}_{N_s})}$, and ${\cal
L}_0$ and ${\cal L}_{N_s}$ denote the maximum likelihoods of the
fits without and with the signal component, respectively. 
We vary each calibration constant for the signal PDF by $\pm 1
\sigma$ and obtain systematic errors from the change in the signal
yield. Adding these errors in quadrature, the systematic error from signal PDF 
is $^{+3.0}_{-3.1}\%$. 
 
In order to obtain the branching fraction, we divide the signal
yield by the reconstruction efficiency, measured from MC to be
$12.8\%$, and by the number of $\bb$ pairs. 
We consider systematic errors in the
reconstruction efficiency due to possible differences between data
and MC. We vary the yields of $\rho^{\pm} \pi^0$ and off-time events by 
$\pm 1\sigma$  and obtain the systematic
 error $^{+4.5}_{-4.7}$\% and $^{+3.1}_{-2.3}$\%, respectively.
 We assign a total error of $8\%$ due to $\pi^0$
reconstruction efficiency, measured by comparing the ratio of the
yields of the $\overline{D}^0 \to K^+ \pi^-$ and $\overline{D}^0 \to K^+ \pi^- \pi^0$ decays. The experimental errors on the branching fractions for
these decays~\cite{PDG} are included in this value. We check the
effect of the continuum suppression using a control sample of
$\bp\to\dzbpip$ decays; the ${\cal R}_{\rm sig}$ requirements has a similar
efficiency for the MC control sample and for signal MC. Comparing the
${\cal R}_{\rm s}$ requirement on the control sample in data and MC,
a systematic error of $1.5\%$ is assigned.
Finally, we assign a systematic error of $1.3\%$ due to the
uncertainty in the number of $\bb$ pairs, ($534.6 \pm 7.0$) $\times \
10^6$, and obtain a branching fraction of
\begin{displaymath}
{\cal B}(B^0\rightarrow\pi^0\pi^0)={\BR}.
\end{displaymath}
 
The significance including systematic uncertainties
is reduced to $5.4 \sigma$.
\begin{figure}
\includegraphics[width=0.45\textwidth]{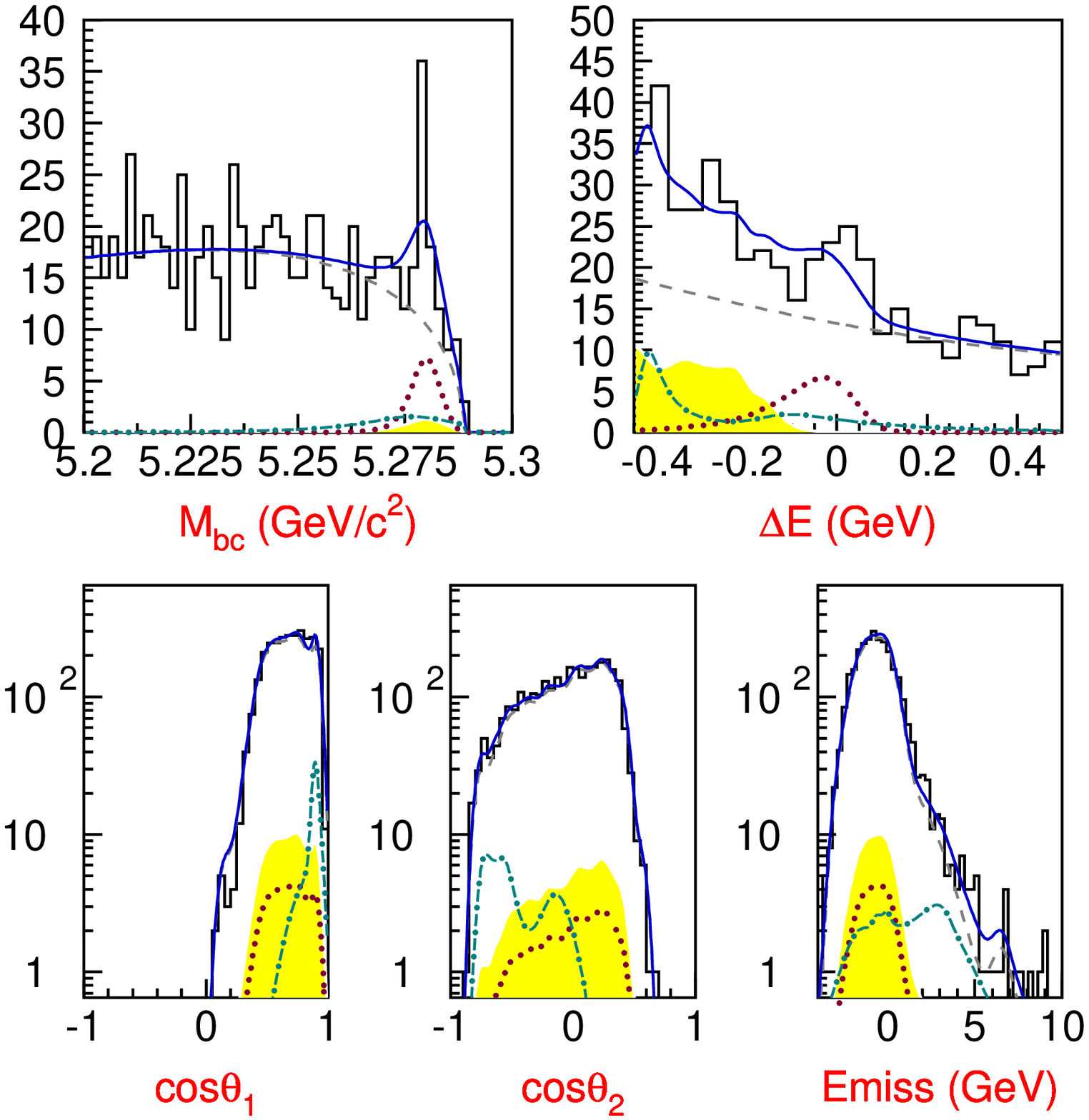}
\includegraphics[width=0.45\textwidth]{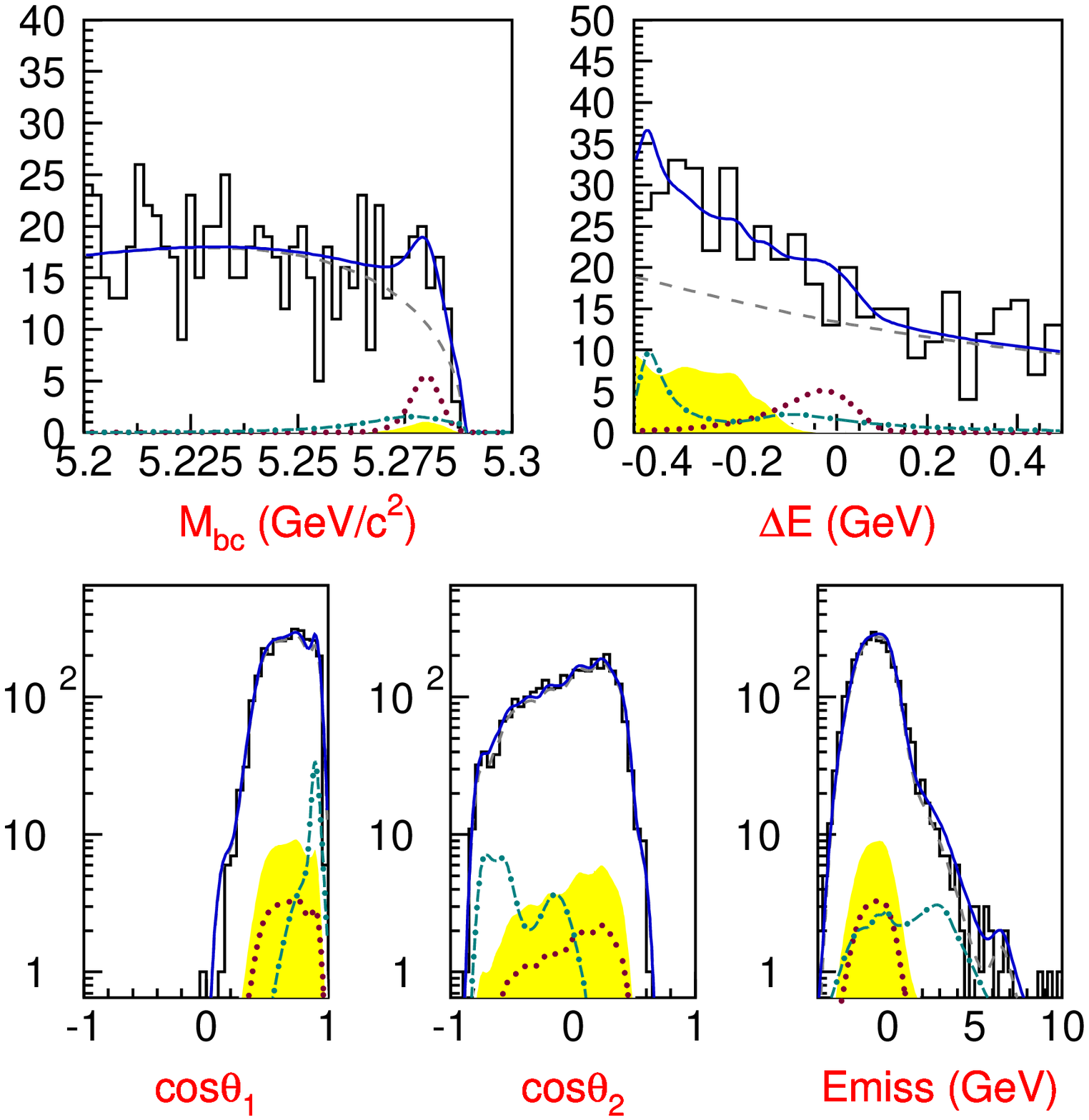}
  \caption{
    \label{fig:tagged}
    $\mbc$, $\de$, $\cos \theta_1$, $\cos \theta_2$ and $E_{\rm miss}$ 
    distributions with projections of the fit
    superimposed. The distributions are shown separately for
    events tagged as $\overline{B}{}^0$ (left) and ${B}^0$ (right).
  }
\end{figure}
 
This new measurement is lower than our previous published result, 
$(2.3^{+0.4+0.2}_{-0.5-0.3})\times 10^{-6}$, in which the QED background
was considered in the systematic uncertainty. We perform a 
$M_{\rm bc}$-$\Delta E$ fit to the 
data used in the previous analysis. Although the current analysis has 
tighter ${\mathcal R}_{\rm sig}$ requirements to suppress the continuum background, the obtained value of the branching fraction is $(2.0^{+0.5}_{-0.4})\times 10^{-6}$, consistent with the previous
result. We then perform a five-dimensional fit to the same data, the 
central value of the branching fraction drops to $(1.7\pm 0.4)\times  10^{-6}$,
which is larger than the measurement with 535 million $B\overline B$ pairs. 
We also compare the 2-d and 5-d fit results using the full dataset, the 
difference of the branching fraction is $0.3 \times 10^{-6}$. Therefore, we
conclude that the change in the branching fraction is due to a statistical 
fluctuation and the inclusion of the off-time QED background in the fit, where 
the former has a larger impact.

The result for ${\cal A}_{CP}$ is $0.44^{+0.73+0.04}_{-0.62-0.06}$.
Systematic errors are estimated by varying the fitting parameters by
$\pm 1\sigma$. Including the errors of wrong-tag fraction ($w_k$) and the
time-integrated mixing parameter ($\chi_d$), 
the total systematic error is $^{+9.6}_{-17.7}$\%.
To illustrate this asymmetry, we show the results separately for
$B^0$ and $\overline{B}{}^0$ tags in Fig.~\ref{fig:tagged}.

In conclusion, we have improved the measurements of  $\bz\to\pizpiz$
in a data sample of 535 million $\bb$ pairs.  We
obtain $74.4^{+21.4}_{-19.7}$ signal events with a significance of
$5.4$ standard deviations ($\sigma$) including
systematic uncertainties. The branching fraction is measured to be
$\BR$. The branching fraction is different from our
previous result due to the larger data sample that is used and the treatment of 
the off-time QED background. We 
also report the direct $CP$ asymmetry to be 0.44$^{+0.73+0.04}_{-0.62-0.06}$,
which is consistent with our previous result \cite{pi0pi0_Belle}.
The branching fraction for
$\bz\to\pizpiz$, together with the measurements of its direct $CP$
violating asymmetry ${\cal A}_{CP}$, will allow a model-independent
extraction of the CKM angle $\phi_2$ from measurements
of the $B\to \pi\pi$ system in the near future.
 

We thank the KEKB group for the excellent operation of the
accelerator, the KEK cryogenics group for the efficient
operation of the solenoid, and the KEK computer group and
the National Institute of Informatics for valuable computing
and Super-SINET network support. We acknowledge support from
the Ministry of Education, Culture, Sports, Science, and
Technology of Japan and the Japan Society for the Promotion
of Science; the Australian Research Council and the
Australian Department of Education, Science and Training;
the National Science Foundation of China and the Knowledge
Innovation Program of the Chinese Academy of Sciences under
contract No.~10575109 and IHEP-U-503; the Department of Science
and Technology of India; the BK21 program of the Ministry of Education of
Korea, and the CHEP SRC program and Basic Research program
(grant No. R01-2005-000-10089-0) of the Korea Science and
Engineering Foundation; the Polish State Committee for
Scientific Research under contract No.~2P03B 01324; the
Ministry of Science and Technology of the Russian
Federation; the Slovenian Research Agency;
the Swiss National Science Foundation; the National Science Council and
the Ministry of Education of Taiwan; and the U.S.\
Department of Energy.


\end{document}